\documentstyle[aps,preprint,epsfig]{revtex}
\begin{document}
\tightenlines
\draft

\title{Radiative mass in QCD at high density}

\author{Deog Ki Hong\footnote{
E-mail address: {\tt dkhong@hyowon.cc.pusan.ac.kr}
}}

\address{
Department of Physics, Pusan National University,
Pusan 609-735, Korea}

\maketitle

\begin{abstract}
We show that radiatively generated Majorana mass for antiquarks
is same as the Cooper-pair gap. We then calculate
the electromagnetic corrections to mass of
particles in the color-flavor locking phase of QCD at high density.
The mass spectrum forms multiplets
under $SU(2)_V\times U(1)_V$. The charged pions and kaons get the
electromagnetic mass which is proportional to the Cooper-pair gap.
\end{abstract}


It has been known for more than two decades
that the ground state of dense matter is color
superconductor~\cite{bailin}.
However, intense study on color superconductivity
has begun quite recently, after realizing that it may lead to
a large superconducting gap~\cite{rapp,arw98},
which can be explored by heavy ion collision experiments or
in the core of neutron stars.
Superconducting quark matter has two different phases, depending
on density. At intermediate density, the Cooper pair is color
anti-triplet but flavor singlet, breaking only the color symmetry
down to a subgroup, $SU(3)_c\mapsto SU(2)_c$.
For high density where the strange quark
is light, compared to the chemical
potential, $m_s<\mu$, the Cooper-pair condensate is predicted
to take a so-called color-flavor locking (CFL) form~\cite{arw99}:
\begin{equation}
\left< \psi_{Li}^{a}(\vec p)\psi_{Lj}^{b}(-\vec p) \right>
=-\left<\psi_{Ri}^{a}(\vec p)\psi_{Rj}^{b}(-\vec p)\right>
=k_1\delta_i^a\delta_j^b+k_2\delta_j^a\delta_i^b,
\label{cfl}
\end{equation}
where $i,j=1,2,3$ and $a,b=1,2,3$ are flavor and color
indices, respectively. At much higher density
($\mu\gg\Lambda_{\rm QCD}$), $k_1(\equiv \Delta_0)\simeq -k_2$
and color-flavor locking phase is shown to be
energetically preferred~\cite{hsu,my2}.

Recently, the meson mass in the CFL phase has been calculated by
several groups~\cite{ss,my3,Rho:2000xf,manuel,savage}. It is found
that the (pseudo) Nambu-Goldstone boson mass due to the current
quark mass vanishes at high density, since the Dirac mass
connects quark with anti-quark and is thus a subleading operator
in $1/\mu$ expansion, provided that the Majorana mass for
anti-quarks, called antigap, is not too large. In this letter, we
calculate the Majorana mass (or antigap) of antiquarks to
implement the above results and then we calculate the
electromagnetic corrections, which is a leading order in $1/\mu$
expansion, to the particle mass in the CFL phase, using the high
density effective theory developed in~\cite{my1,my2}.

At low energy, $E<\mu$, the quarks near the Fermi surface are relevant
degrees of freedom while anti-quarks, the holes deep in the Dirac sea,
are irrelevant and decoupled, since
the energy of quarks near the Fermi surface in perturbative QCD
is $E_+=-\mu+\sqrt{{\vec p}^2+m^2}\simeq \vec l\cdot \vec v_F<\mu$,
while anti-quarks have energy $E_-=\mu+\sqrt{{\vec p}^2+m^2}
\simeq 2\mu+\vec l\cdot \vec v_F>\mu$, where
we decompose the quark momentum as $\vec p=\vec p_F+\vec l$
with $|\vec l|<\mu$, and
$\vec v_F\equiv\vec p_F/\mu$ is the Fermi velocity.

Since
an arbitrarily small energy can create a pair of quarks and holes
near the Fermi surface, quarks and holes can form pairs and condense
to gain energy if any attraction is provided. It turns out that
color anti-triplet diquark condensates are energetically more
preferred, compared to particle-hole condensates~\cite{shson}.
Since magnetic gluons are not screened in dense quark matter,
the (long-range) magnetic gluon exchange interaction leads
to a much bigger Cooper-pair gap,
which is in hard-dense loop (HDL) approximation given
as~\cite{son1,my4,sw,pr,hs},
\begin{equation}
\Delta_0
\sim {\mu\over g_s^5}\exp\left(-{3\pi^2\over\sqrt{2}g_s}\right).
\label{gap}
\end{equation}

Being irrelevant modes and thus decoupled from the low energy
dynamics, anti-quarks do not get dynamical Majorana mass
(namely, do not form condensates) in dense matter,
since it is energetically not preferred. However,
they will get a Majorana mass radiatively
once a Cooper-pair gap opens for quarks (or holes) near the Fermi surface,
since all symmetries that forbid Majorana mass for anti-quarks are
then broken.
The Cooper-pair gap, $\Delta$,
can be transmitted to anti-quarks through
their couplings to quarks near the Fermi surface or to gluons.
In the leading order in loop expansion, there are two diagrams that
contribute to the anti-quark Majorana mass, denoted as $\bar\Delta_0$,
shown in Fig.~1.
From the first diagram [Fig.~1(a)], we get
\begin{eqnarray}
\bar\Delta_0=(-ig_s)^2\int{d^4l\over(2\pi)^4}
\gamma^{\mu}_{\perp}{T^A\Delta T^A\over l_{\parallel}^2-
\Delta^2(l_{\parallel})}\gamma_{\perp}^{\nu}D_{\mu\nu}(l),
\label{antigap}
\end{eqnarray}
where $\gamma_{\perp}^{\mu}=\gamma^{\mu}-\gamma^{\mu}_{\parallel}$
with
$\gamma_{\parallel}^{\mu}=(\gamma^0,\vec v_F\vec\gamma\cdot\vec v_F)$
and $D_{\mu\nu}$ is the (Higgsed) gluon propagator.
Note that antiquarks
interact with quarks near the Fermi surface via the magnetic gluons only,
whose propagator is given Euclidean space as~\cite{my4}
\begin{equation}
D_{\mu\nu}^{m}(l)={P_{\mu\nu}^T}{|\vec l|\over
|\vec l|^3+M_0^2\Delta_0+\pi M^2|l_0|/2},
\end{equation}
where the polarization tensor
$P^T_{ij}=\delta_{ij}-l_il_j/|\vec l|^2$ and $P^T_{00}=0=P_{0i}^T$,
the Meissner mass $M_0^2=\alpha_s\mu^2/2$ and the Debye screening
mass $M^2=3g_s^2\mu^2/(4\pi^2)$. Taking the trace over gamma matrices in
Eq.~(\ref{antigap}), we get in Euclidean space
\begin{eqnarray}
\bar\Delta_0={2g_s^2\over3}\int{{\rm d}^4l\over(2\pi)^4}
{|\vec l|\over
|\vec l|^3+M_0^2\Delta_0+\pi M^2|l_0|/2}
\cdot {\Delta(l_{\parallel})\over l_{\parallel}^2+\Delta^2},
\end{eqnarray}
which is nothing but the Cooper-pair gap at zero
external momentum except that here we have
Higgsed magnetic gluons only, instead of full gluons.
If we take $\Delta(l_{\parallel})\simeq \Delta_0$,
we get
\begin{equation}
\bar\Delta_0=\Delta_0\cdot {g_s^2\over 72\pi^2}\ln\left({4\mu^2\over
\Delta_0^2}\right)\cdot \ln\left({48\mu^3\over 5\pi M^2\Delta_0}\right)
=\Delta_0\left(1+O(g_s)\right),
\end{equation}
where we used in the logarithm
$\Delta_0\sim \mu g_s^{-5}\exp\left(-6\pi/g_s\right)$,
the solution to the gap equation in the constant gap
approximation~\cite{my4,pr}.
The contribution from the second diagram in Fig.~1(b) is
$\sim g_s^2\Delta_0\ln(\mu^2/\Delta_0^2)$, which is
suppressed by $O(g_s)$. Therefore, we find that in the leading order
the Majorana mass for the anti-quarks is same as that of quarks and holes
near the Fermi surface. Because of this Majorana mass term of antiquarks,
additional corrections will be generated to the Dirac mass term.
As in~\cite{my1,my2}, if we introduce the charge conjugated
field $\psi_c=C{\bar\psi}^T$ with $C=i\gamma^0\gamma^2$
and decompose the quark field into
states ($\psi_+$) near the Fermi surface and the states
($\psi_-$) deep in the Dirac sea, the
Dirac mass term can be rewritten as
\begin{equation}
m_q\bar\psi\psi={1\over2}m_q
\left(\bar\psi_+\psi_-+\bar\psi_-\psi_+\right)+
{1\over2}m_q^T
\left(\bar\psi_{c+}\psi_{c-}+\bar\psi_{c-}\psi_{c+}\right).
\end{equation}
Since now the states in the Dirac sea can propagate into their
charge conjugated states via the radiatively generated
Majorana mass term,
the Dirac mass term becomes, if one integrates out $\psi_-$ fields,
\begin{equation}
m_q\bar\psi\psi={m_q^2\over2\mu}\psi_+^{\dagger}\left(
1-{i\partial\cdot V\over2\mu}\right)\psi_+
+{m_qm_q^T\over4\mu^2}\psi_+^{\dagger}\Delta_0\psi_{c+}+\cdots,
\end{equation}
where $V^{\mu}=(1,\vec v_F)$ and the ellipsis denotes the terms
higher order in $1/\mu$. Then, the vacuum energy shift due to the
Dirac mass term is $\sim m_q^2\Delta_0^2\ln(\mu^2/\Delta_0^2)$ in
the leading order,\footnote{In the Erratum~\cite{ss}, Son and Stephanov
also found the $1/\mu$ dependence of the meson mass. But, they claim
that there is no logarithmic correction to the meson mass, contrary to
all others' results~\cite{my3,Rho:2000xf,manuel,savage}. In the effective
theory point of view, the operator, which corresponds to
the cross diagram they claim to cancel the logarithm,
does not exist in the leading order.}
which shows the meson mass due to the Dirac
mass indeed vanishes at asymptotic density, $m_{\pi}^2\sim m_q^2
\Delta_0^2/\mu^2\cdot \ln(\mu^2/\Delta_0^2)$, since the vacuum
energy in the meson Lagrangian is $m_{\pi}^2F^2$ with the pion
decay constant $F\sim\mu$~\cite{ss,Rho:2000xf}.


Now, we calculate the electromagnetic mass of particles in the CFL phase.
Since the color and flavor are locked in the CFL phase,
simultaneous rotations of color and flavor space are only unbroken,
leading to an interesting symmetry-breaking pattern,
\begin{equation}
SU(3)_c\times SU(3)_L\times SU(3)_R\times U(1)_B
\mapsto SU(3)_V\times Z_2,
\end{equation}
where $SU(3)_V$ is the diagonal subgroup of three $SU(3)$'s.
The particle spectrum of the CFL phase is then as following:
There are 8 massive gluons of mass, $m_g\sim g_s\mu$ by
Higgs mechanism, which form an vector meson octet under $SU(3)_V$
and there are 9 massive quarks, $\psi^a_i$, which are octet and
singlet under $SU(3)_V$. The octet, $\psi_8=P_8\psi$, has mass
$k_2$ and the singlet, $\psi_S=P_1\psi$, has mass $3k_1+k_2$,
which can be seen if we project out the gap as
\begin{equation}
k_1\delta_i^a\delta_j^b+k_2\delta_j^a\delta_i^a
=k_2Q_{8ij}^{ab}+(3k_1+k_2)P_{1ij}^{ab}
\end{equation}
where the projectors are
$Q_{8ij}^{ab}=\delta_j^a\delta_i^b-1/3\delta_i^a\delta_j^b$,
$P_{1ij}^{ab}=1/3\delta_i^a\delta_j^b$,
and $Q_8^2=P_8$~\cite{zarembo}.
Finally, we have nine Nambu-Goldstone (NG) bosons, consisting of
meson octet and singlet, associated with chiral symmetry breaking
and baryon superfluidity, respectively.

Since the Cooper-pair carries electric charge, there will be mixing
between gluons and photons through the Cooper-pair gap.
A linear combination of two fields, called a modified photon,
$\tilde A_{\mu}=A_{\mu}\cos\theta+A_{\mu}^Y\sin\theta$,
remain massless and couples to fields with
strength $\tilde e=e\cos\theta$,
where $\tan\theta=e/g_s$ and $A_{\mu}^Y$ is the gluon
field corresponding to the color
hypercharge, $Y={\rm diag} (-2/3,1/3,1/3)$.
Once we turn on the electromagnetism, the global symmetry
$SU(3)_V$ will break explicitly down to $SU(2)_V\times U(1)_V$, since
the electric charge $Q={\rm diag.}(2/3,-1/3,-1/3)$ is not flavor
singlet or equivalently the charge of $U(1)_{\tilde Q}$,
modified electromagnetism is not identity but given
in the color-flavor locked space as
\begin{equation}
{\tilde Q_{ai}}=\left(\begin{array}{c c c}
                      0&-1&-1 \\
                      1&0&0 \\
                      1&0&0
                 \end{array}  \right).
\end{equation}
Under $SU(2)_V$, the octets of $SU(3)_V$ are decomposed as
$8=3\oplus 2\oplus\bar2\oplus1$, among which the doublet has
$\tilde Q=+1$ and the anti-doublet has $\tilde Q=-1$ and
all others together with $SU(3)_V$ singlets
are neutral under $U(1)_{\tilde Q}$.
All particles with $U(1)_{\tilde Q}$ charge will get
(modified) electromagnetic
masses.

We first calculate the electromagnetic mass to the mesons.
The leading-order contribution to the vacuum energy, due to the
(modified) electromagnetic interaction, is given by a diagram,
shown in Fig.~2, which is
\begin{equation}
\delta E=
-{{\tilde e}^2\over 2}\sum_A\int {\rm d}^4x
{\tilde D}^{\mu\nu}(x)
\left<\Omega\right|{\rm T}J_{\mu}^A(x)J_{\nu}^A(0)\left|
\Omega\right>,
\label{vac}
\end{equation}
where $\left|\Omega\right>$ denotes the CFL vacuum and
${\tilde D}_{\mu\nu}$ is the modified photon propagator
in HDL approximation. At low energy, modified photons
will not interact with color superconductors. But at high energy,
modified photons
of energy larger than twice of the gap
will get screened and Landau damped,
with a screening mass
${\tilde M}^2\simeq(8/9)\sin^2\theta M^2$,
since it can break Cooper-pairs and scatter
with the gaped quarks in the Fermi sea,

Since the main contributions to the loop integration in
Eq.~(\ref{vac}) come from photon energy larger than the gap,
we get
\begin{equation}
V_{\rm em}=4{{\tilde e}^2\over8\pi}{\Delta_0}^2\mu^2
\ln\left({\mu^2\over \Delta_0^2}\right)\cdot \left[\ln
\left({16\mu^3\over \pi{\tilde M}^2\Delta_0}\right)
\right]^2\simeq18\pi \left({\tilde e\over g_s}\right)^2\Delta_0^2\mu^2
\ln\left({\mu^2\over\Delta_0^2}\right),
\end{equation}
where the gap in the second logarithm is traded with the coupling $g_s$
by Eq.~(\ref{gap}).
The mass of (pseudo) Nambu-Goldstone bosons is then just
the curvature of the vacuum energy at the origin of
the NG boson manifold~\cite{peskin},
\begin{equation}
m^2_{ab}={1\over F^2}{\partial^2\over \partial\alpha_a
\partial\alpha_b}\delta E(U)|_{U=1},
\end{equation}
where $U$ is a unitary matrix in the chiral group and
$\alpha_a$'s are its coordinates in the group space.
We therefore find the charged
NG bosons get the (modified) electromagnetic mass
\begin{equation}
m^2_{\pi}={9\pi\over4} \left({\tilde e\over g_s}\right)^2
\Delta_0^2{\mu^2\over F^2}\ln\left({\mu^2\over\Delta_0^2}\right).
\end{equation}
Since the pion decay constant $F\simeq 0.209\mu$~\cite{ss},
the charged NG bosons get
the electromagnetic mass
$m_{\pi}\simeq 12.7\sin\theta\Delta_0
\left[\ln(\mu^2/\Delta_0^2)\right]^{1/2}$.

Finally, we calculate the (modified) electromagnetic mass
of charged gluons and quarks. Since the (modified) photon
contributes to the self energy of charged gluons and quarks,
we just calculate the one-loop corrections due to the (modified)
photon to the self energy to get
electromagnetic mass for the charged gluons and quarks
\begin{equation}
\delta m_g^2\sim{\tilde e}^2\Delta_0^2
\ln\left({\mu^2\over\Delta_0^2}\right)
\quad{\rm and}\quad
\delta m_{\psi}\sim{\tilde e}^2\Delta_0
\ln\left({\mu^2\over\Delta_0^2}\right).
\end{equation}

To conclude, we show that the anti-quarks, holes deep in the
Dirac sea, get Majorana mass radiatively, which is same as the
Cooper-pair gap. We calculate the (modified) electromagnetic mass
of particles in the CFL color superconductor. As the modified
electromagnetic interaction breaks explicitly the global
$SU(3)_V$ symmetry down to $SU(2)_V\times U(1)_V$, the degeneracy
in the mass spectrum is lifted such that among octets the
$SU(2)_V$ doublets become more massive. Especially, the charged
pions and kaons get electromagnetic mass, of order of the
Cooper-pair gap, and do not vanish at high density, while neutral
mesons become massless at asymptotic density.


\acknowledgments

The author wishes to thank T. Lee, D.-P. Min, and K. Rajagopal
for useful comments and the Institute for Nuclear Theory at the
University of Washington, where part of this work was done,
for its hospitality.
This work was supported  by the academic research fund of Ministry of
Education, Republic of Korea, Project No. BSRI-99-015-DI0114.

\newpage
\begin{figure}
\vskip 0.2in
\centerline{\epsfbox{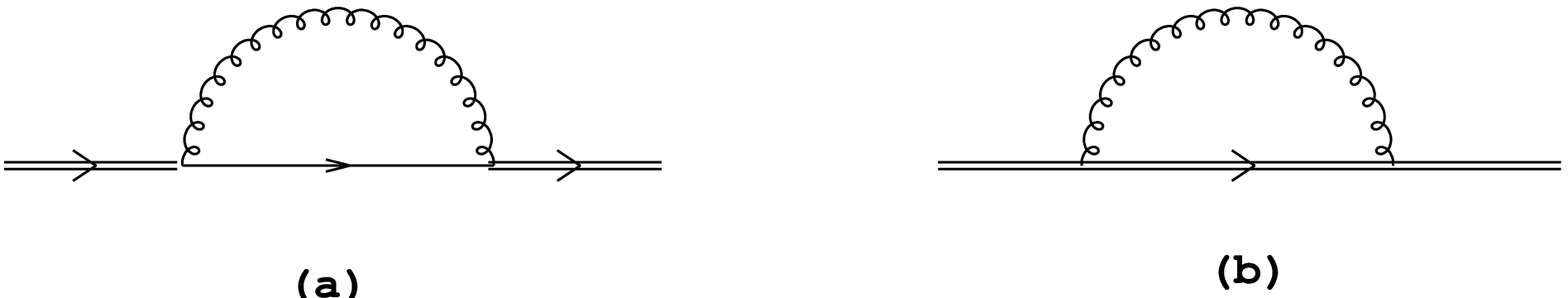}}
\vskip 0.2in
\caption{The radiative correction to the Majorana mass of antiquarks.
The double solid line denotes antiquarks and the single solid
line denotes the quarks near the Fermi surface.}
\end{figure}
\vskip 0.2in
\begin{figure}
\centerline{\epsfbox{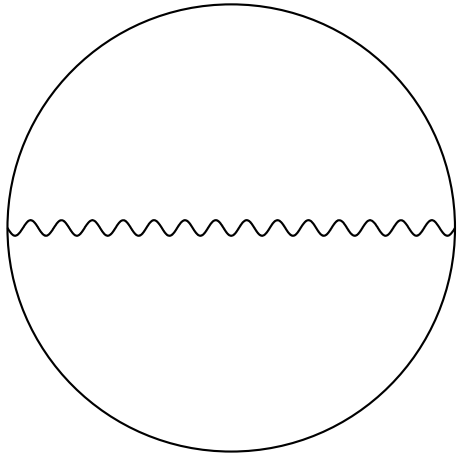}}
\vskip 0.2in
\caption{The (modified) electromagnetic contribution to the vacuum
energy in the leading order.
The wiggly line denotes the modified photon in the HDL approximation.
}
\end{figure}


\begin{references}

\bibitem{bailin}F. Barrois,
Nucl. Phys. B {\bf 129}, 390 (1977);
D. Bailin and A. Love, Phys. Rep. {\bf 107},
325 (1984); S.~C. Frautschi,
{\it Asymptotic freedom and color superconductivity in dense
quark matter}, in: Proceedings of the
Workshop on Hadronic Matter at Extreme Energy Density, N. Cabibbo, Editor,
Erice, Italy (1978).


\bibitem{rapp}R. Rapp, T. Schafer, E.V. Shuryak, and M. Velkovsky,
Phys. Rev. Lett. {\bf 81}, 53 (1998).

\bibitem{arw98}M. Alford, K. Rajagopal, and F. Wilczek,
Phys. Lett. B {\bf 422}, 247 (1998), hep-ph/9711395.

\bibitem{arw99}M. Alford, K. Rajagopal, and F. Wilczek,
Nucl. Phys. B {\bf 538}, 443 (1999),
hep-ph/98044233.

\bibitem{hsu}
N.~Evans, J.~Hormuzdiar, S.~D.~Hsu and M.~Schwetz,
Nucl.\ Phys.\  {\bf B581}, 391 (2000), hep-ph/9910313;
T.~Schafer,
Nucl.\ Phys.\  {\bf B575}, 269 (2000), hep-ph/9909574.

\bibitem{my2}
D.~K.~Hong,
Nucl.\ Phys.\  {\bf B582}, 451 (2000), hep-ph/9905523.

\bibitem{ss}D.~T. Son and M.~A. Stephanov, Phys. Rev. D {\bf 61},
074012 (2000), hep-ph/9910491;
Erratum, {\bf D62}, 059902 (2000), hep-ph/0004095.

\bibitem{my3}D.~K. Hong, T. Lee, and D.-P. Min, Phys. Lett. B
{\bf 477}, 137 (2000), hep-ph/9912531.

\bibitem{Rho:2000xf}
M.~Rho, A.~Wirzba and I.~Zahed,
Phys.\ Lett.\  {\bf B473}, 126 (2000). 

\bibitem{manuel}C. Manuel and M.~H. Tytgat, Phys. Lett. B {\bf 479},
190 (2000), hep-ph/0001095.

\bibitem{savage}
S.~R.~Beane, P.~F.~Bedaque and M.~J.~Savage,
Phys.\ Lett.\  {\bf B483}, 131 (2000). hep-ph/0002209.

\bibitem{my1}D.~K. Hong, Phys. Lett. B {\bf 473}, 118 (2000),
hep-ph/9812510.

\bibitem{shson}
E.~Shuster and D.~T.~Son,
Nucl.\ Phys.\  {\bf B573}, 434 (2000), hep-ph/9905448;
B.~Park, M.~Rho, A.~Wirzba and I.~Zahed,
Phys.\ Rev.\  {\bf D62}, 034015 (2000), hep-ph/9910347.

\bibitem{son1}
D.~T.~Son,
Phys.\ Rev.\  {\bf D59}, 094019 (1999),
hep-ph/9812287.


\bibitem{my4}D.~K.~Hong, I.~Shovkovy, V.~Miransky, and
L.~C.~R.~Wijewardhana, Phys. Rev. D {\bf 61}, 056001 (2000).


\bibitem{sw}
T.~Sch\"afer and F.~Wilczek,
Phys. Rev. D {\bf 60}, 114033 (1999), hep-ph/9906512;

\bibitem{pr}
R.~Pisarski and D.~Rischke,
     Phys. Rev. D {\bf 61}, 074017 (2000), nucl-th/9910056;

\bibitem{hs}
W.~E.~Brown, J.~T.~Liu and H.~Ren,
Phys.\ Rev.\  {\bf D61}, 114012 (2000), hep-ph/9908248;
S.~D.~Hsu and M.~Schwetz,
Nucl.\ Phys.\  {\bf B572}, 211 (2000), hep-ph/9908310.


\bibitem{zarembo}
K.~Zarembo,
Phys.\ Rev.\  {\bf D62}, 054003 (2000), hep-ph/0002123.


\bibitem{peskin}M.~E. Peskin, Nucl. Phys. B {\bf 175}, 197 (1980);
J. Preskill, {\it ibid.} {\bf 177}, 21 (1981).

\end{references}
\end{document}